# Orbital Period Change of Dimorphos Due to the DART Kinetic Impact

Cristina A. Thomas[1], Shantanu P. Naidu[2], Peter Scheirich[3], Nicholas A. Moskovitz[4], Petr Pravec[3], Steven R. Chesley[2], Andrew S. Rivkin[5], David J. Osip[6], Tim A. Lister[7], Lance A. M. Benner[2], Marina Brozović[2], Carlos Contreras[6], Nidia Morrell[6], Agata Rożek[8], Peter Kušnirák[3], Kamil Hornoch[3], Declan Mages[2], Patrick A. Taylor[9], Andrew D. Seymour[10], Colin Snodgrass[8], Uffe G. Jørgensen[11], Martin Dominik[12], Brian Skiff[4], Tom Polakis[4], Matthew M. Knight[13], Tony L. Farnham[14], Jon D. Giorgini[2], Brian Rush[2], Julie Bellerose[2], Pedro Salas[10], William P. Armentrout[10], Galen Watts[10], Michael W. Busch[15], Joseph Chatelain[7], Edward Gomez[7,16], Sarah Greenstreet[17], Liz Phillips[18,7], Mariangela Bonavita[8], Martin J. Burgdorf[19], Elahe Khalouei[20], Penélope Longa-Peña[21], Markus Rabus[22], Sedighe Sajadian[23], Nancy L. Chabot[5], Andrew F. Cheng[5], William H. Ryan[24], Eileen V. Ryan[24], Carrie E. Holt[14], Harrison F. Agrusa[14]

[1] Northern Arizona University, Flagstaff, AZ, USA
[2] Jet Propulsion Laboratory, California Institute of Technology, Pasadena, CA, USA
[3] Astronomical Institute of the Czech Academy of Sciences, Ondřejov, Czech Republic
[4] Lowell Observatory, Flagstaff, AZ, USA
[5] Johns Hopkins University Applied Physics Laboratory, Laurel, MD, USA
[6] Carnegie Institution for Science, Las Campanas Observatory, La Serena, Chile
[7] Las Cumbres Observatory, Goleta, CA, USA
[8] University of Edinburgh, Royal Observatory, Edinburgh, UK
[9] National Radio Astronomy Observatory, Charlottesville, VA, USA
[10] Green Bank Observatory, Green Bank, WV, USA
[11] Niels Bohr Institute, University of Copenhagen, Copenhagen, Denmark
[12] University of St Andrews, St Andrews, UK
[13] United States Naval Academy, Annapolis, MD, USA
[14] University of Maryland, College Park, MD, USA
[15] SETI Institute, Mountain View, CA, USA
[16] Cardiff University, Cardiff, UK
[17] University of Washington, Seattle, WA, USA
[18] University of California, Santa Barbara, CA, USA
[19] Universität Hamburg, Hamburg, Germany
[20] Seoul National University, Gwanak-gu, Seoul, Korea
[21] Universidad de Antofagasta, Antofagasta, Chile
[22] Universidad Católica de la Santísima Concepción
[23] Isfahan University of Technology, Isfahan, Iran
[24] Magdalena Ridge Observatory, New Mexico Institute of Mining and Technology, Socorro, NM, USA

**The Double Asteroid Redirection Test (DART) spacecraft successfully performed the first test of a kinetic impactor for asteroid deflection by impacting Dimorphos, the secondary of near-Earth binary asteroid (65803) Didymos, and changing the orbital period of Dimorphos. A change in orbital period of approximately 7 minutes was expected if the incident momentum from the DART spacecraft was directly transferred to the asteroid target in a perfectly inelastic collision[1], but studies of the probable impact conditions and asteroid properties indicated that a considerable momentum enhancement ($\boldsymbol{\beta}$) was possible[2,3]. In the years prior to impact, we used lightcurve observations to accurately determine the pre-impact orbit parameters of Dimorphos with respect to Didymos[4–6]. Here we report the change in the orbital period of Dimorphos as a result of the DART kinetic impact to be -33.0 ± 1.0 (3$\sigma$) minutes. Using new Earth-based lightcurve and radar observations, two independent approaches determined identical values for the change in the orbital period. This large orbit period change suggests that ejecta contributed a significant amount of momentum to the asteroid beyond what the DART spacecraft carried.**

NASA's DART (Double Asteroid Redirection Test) successfully impacted Dimorphos, the secondary of the near-Earth binary asteroid (65803) Didymos, on 26 September 2022 at 23:14 UTC. The primary objective of DART was to change the orbital period of Dimorphos around Didymos to demonstrate that a kinetic impactor is a viable method of asteroid deflection[1,7]. The mission targeted the secondary asteroid in an eclipsing binary system since the experiment could use a single impacting spacecraft and measure the change in the secondary's orbit through ground-based observations. The Didymos system was selected as the target because it is among the most accessible (low $\Delta$V) of the near-Earth binaries, it has been extremely well characterized[4-6,8–12], and Dimorphos is in the size range identified as relevant for deflection by a kinetic impactor[13,14].

The DART spacecraft collided head-on into the leading hemisphere of Dimorphos in order to maximize the momentum transfer and reduce the semi-major axis of the Dimorphos orbit, resulting in a shorter orbital period[7]. If the incident momentum from the impacting spacecraft was simply transferred to the asteroid target with no additional momentum enhancement, an orbital period change for Dimorphos of roughly seven minutes was expected[1]. Impact simulations conducted in preparation for DART's kinetic impact test indicated that depending on the material strength, impact conditions, and other properties the value of the momentum enhancement factor, $\boldsymbol{\beta}$, could be considerable, with predicted values as high as five[2] or six[3] with a resulting orbital period change of over 40 minutes[15].

The Didymos system lightcurve is composed of three parts: the rotational lightcurve of Didymos, the rotational lightcurve of Dimorphos, and the mutual events that constrain the orbital period. The Didymos rotational lightcurve can be clearly distinguished because the primary contributes approximately 96% of the light from the system. The Dimorphos rotational period has not been resolved due to its comparatively small size, the oblate shape of Dimorphos[16], and the accuracy of the photometric observations necessary for such a detection. Mutual events cause a measurable decrease in the total brightness of the system. We define the primary/secondary occultation or eclipse based on which object is being obscured or shadowed, respectively. We use the timings of the observed mutual events in the determination of the orbital period. For the Didymos-Dimorphos system, mutual events occur when the Didymos-Sun or the Didymos-Earth vector forms an angle less than ~17 degrees with the mutual orbit plane of the system. Since the inclination of the mutual orbit to the heliocentric orbit of the binary system is lower than this value, eclipses (mutual shadowing of the components, Figure 1) always occur. Occultations did not occur during the observing period presented in this paper.

A precise determination of the Didymos system's pre-impact orbital parameters was a key goal once the system was chosen as the target of DART. The initial orbit of Dimorphos was first defined following the 2003 apparition when the secondary was discovered[11,17]. Analyses of lightcurve derived mutual events obtained during 2003-2022[4] led to independent and consistent orbital periods[5,6]. The data used in the published pre-impact orbit solutions were augmented with additional photometric data obtained in July 2022 to calculate the pre-impact orbit period for Dimorphos (Extended Data Table 1). Both approaches determined a statistically identical pre-impact orbital period of 11.92148 ± 0.00013 h (3σ).

To determine the post-impact orbital period, we obtained radar and lightcurve observations of the Didymos system. Our radar observations of Didymos and Dimorphos began about 11 hours after impact using the Goldstone X-band (3.5 cm, 8560 MHz) and continued for 14 dates between UTC 27 September - 13 October (all subsequent dates are in UTC). We also used the Green Bank Telescope to receive radar echoes in a bistatic configuration with transmissions from Goldstone on 2, 6, and 9 October. We obtained echo power spectra during each of the observing windows and range-Doppler images (Figure 2) on ten days centered on 4 October, when the signal to noise ratios (SNRs) were the highest because Didymos was the closest to Earth. The radar observations of the system are not subject to the same shadowing geometry as the lightcurve photometry. Dimorphos can be seen when illuminated by radar and the system was never in a radar eclipse geometry. We measured the separations between Dimorphos and Didymos in the echo power spectra and the range-Doppler images. We used these measurements in the determination of the orbital parameters of Dimorphos relative to Didymos. We only used data in which the SNRs were strong enough to detect both Didymos and Dimorphos. The first observation of Dimorphos ($8\sigma$ detection), approximately 12 hours after impact, yielded the first estimate of the orbital period change of -36 $\pm$ 15 minutes.

Following the DART kinetic impact, ejecta was introduced into the system[18]. The additional flux and the variable brightness from the rapidly evolving ejecta prevented immediate observations of the mutual events. Lightcurve observations began in the hours after impact and our first successful detection of a mutual event was a secondary eclipse approximately 29.5 hours after impact (mid-time at geocentric UT 28 September 04:50). At the time of the first mutual event detection, the flux from the ejecta dominated the signal within the photometric aperture. This contamination resulted in a reduction in the observed amplitude of the Didymos rotational lightcurve by a factor of 3. The apparent depth of the secondary eclipse was also significantly reduced compared to the predictions[6]. Pre-impact ejecta models[19] suggested that it could take up to several days for our ground-based lightcurve observations to detect the first mutual event due to the total ejecta brightness and that the rate of change of that brightness could be comparable to the expected changes in the Didymos system brightness during mutual events.

Photometric observations included in this analysis were obtained from 28 September to 10 October 2022 (Extended Data Table 2). This set of observations ends on 10 October because subsequent observations did not have the required precision due to the bright Moon. On average our data have photometric accuracy of RMS ~0.006 magnitudes. The exceptional quality of the data included in our analysis has enabled the determination of the Dimorphos orbital period change via lightcurves despite the presence of ejecta in all of our observations (Figures 3 & 4). At the time of these first observations, the primary eclipses were grazing events (Figure 1), which required exceptionally precise data to measure.

Two independent methods were used to model the available data for determination of the post-impact orbital period: (1) we use the processes described in ref. [6] to model the lightcurve observations alone and (2) we combine the radar and mutual event timings[5,11] plus Didymos-relative astrometry of Dimorphos in optical navigation images from the DART spacecraft DRACO camera[20]. Both methods use the same ground-based photometric datasets, but have independent processes for accepting individual data points and mutual events. Ellipsoidal approximations of the shapes of Dimorphos and Didymos are incorporated in the calculation of the orbit period of Dimorphos in both approaches and the axial ratios reported in ref. [16] were used for their calculation.

We determine a post-impact period of 11.372 ± 0.017 (3-$\sigma$) hours with a period change of -33.0 ± 1.0 (3-$\sigma$) minutes. Both methods provide statistically identical results for the post-impact orbital period. The rotation period of Didymos is measured during the lightcurve analysis process and shows no variation from its pre-impact value of 2.260 hours to an uncertainty of approximately 5 seconds (3-$\sigma$). The rotational lightcurve of Dimorphos has not been detected. The new orbital period results in Dimorphos completing an additional full orbit every ~9.8 days.

The difference between the pre-impact and post-impact mutual orbit period of the Didymos-Dimorphos system greatly exceeds the ~7 minute period change calculated for the case of a simple momentum transfer with no momentum enhancement[1]. Estimates of the change in orbital velocity imparted to Dimorphos require modeling beyond the scope of this paper, but it is evident that the ejecta from the DART impact carried a significant amount of momentum compared to what the DART spacecraft itself was carrying (e.g., [21]). To serve as a proof-of-concept for the kinetic impactor technique of planetary defense, DART needed to demonstrate that an asteroid could be targeted during a high-speed encounter[16] and that the target's orbit could be changed. DART has successfully done both.

Figure 1:

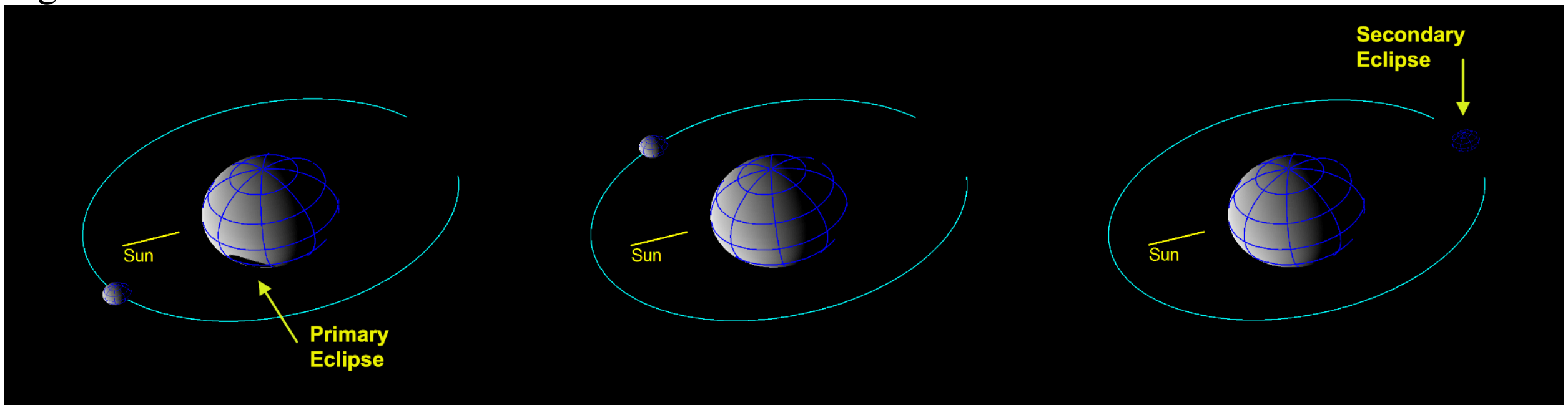

We determine the new orbital period of Dimorphos using the times of mutual events when a measurable decrease in the system brightness occurs due to an eclipse or occultation. Due to the geometry of the Didymos system during this time period, our lightcurve observations include primary eclipses (left), time outside mutual events (center), and secondary eclipses (right). These diagrams simulate the view of the system from Earth on 10 October 06:09 (primary eclipse), 10 October 08:47 (outside events), 10 October 12:06 (secondary eclipse) in geocentric UTC. The primary eclipses observed throughout our post-impact dataset are grazing, which resulted in a subtle decrease in system brightness (Figure 3). During the secondary eclipse, Dimorphos is completely shadowed.

Figure 2:

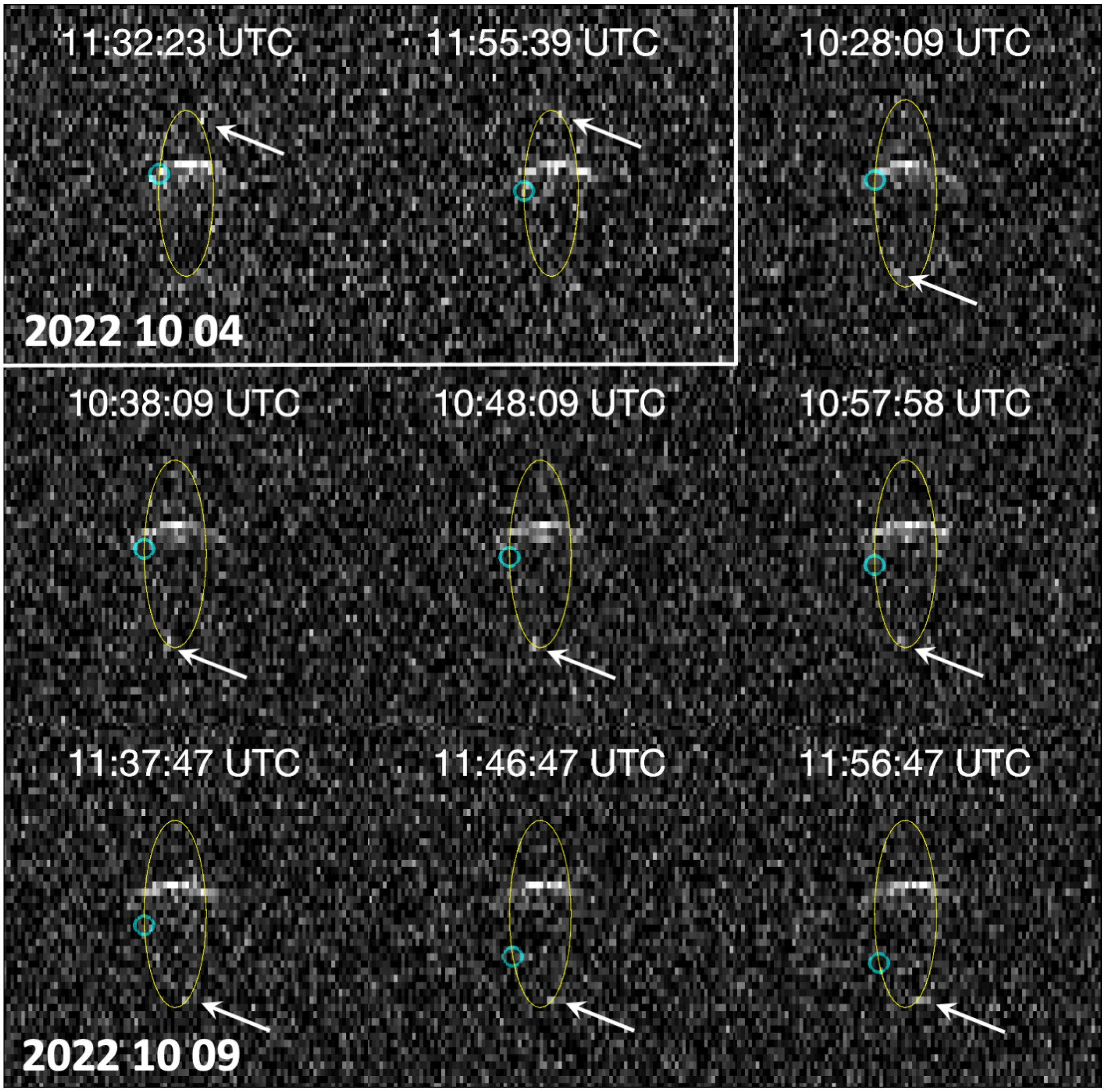

Radar range-Doppler images obtained on October 4 using Goldstone and October 9 using Goldstone to transmit and the Green Bank Telescope to receive. Within each image, distance from Earth increases from top to bottom and Doppler frequency increases to the right, so rotation and orbital motion are counterclockwise. Each image was integrated for 20 minutes, with 10 minutes of overlap between successive images. Images have resolutions of 75 m x 0.5 Hz. The broader echo is from Didymos and the smaller, fainter echo shown using arrows is from Dimorphos. Open circles show Dimorphos positions predicted by the pre-impact orbit. The yellow ellipses show the trajectory of Dimorphos. Prediction uncertainties are smaller than the image resolution. On October 4, the ellipse spans -870 m to +870 m along the y-axis and -7 Hz to +7 Hz along the x-axis, corresponding to line of sight velocity of -12 cm/s to +12 cm/s. On October 9, the ellipse spans -980 m to +980 m along the y-axis and -8 Hz to +8 Hz along the x-axis, corresponding to line of sight velocity of -14 cm/s to +14 cm/s. The physical extents of the ellipse vary due to the viewing geometry.

Figure 3:

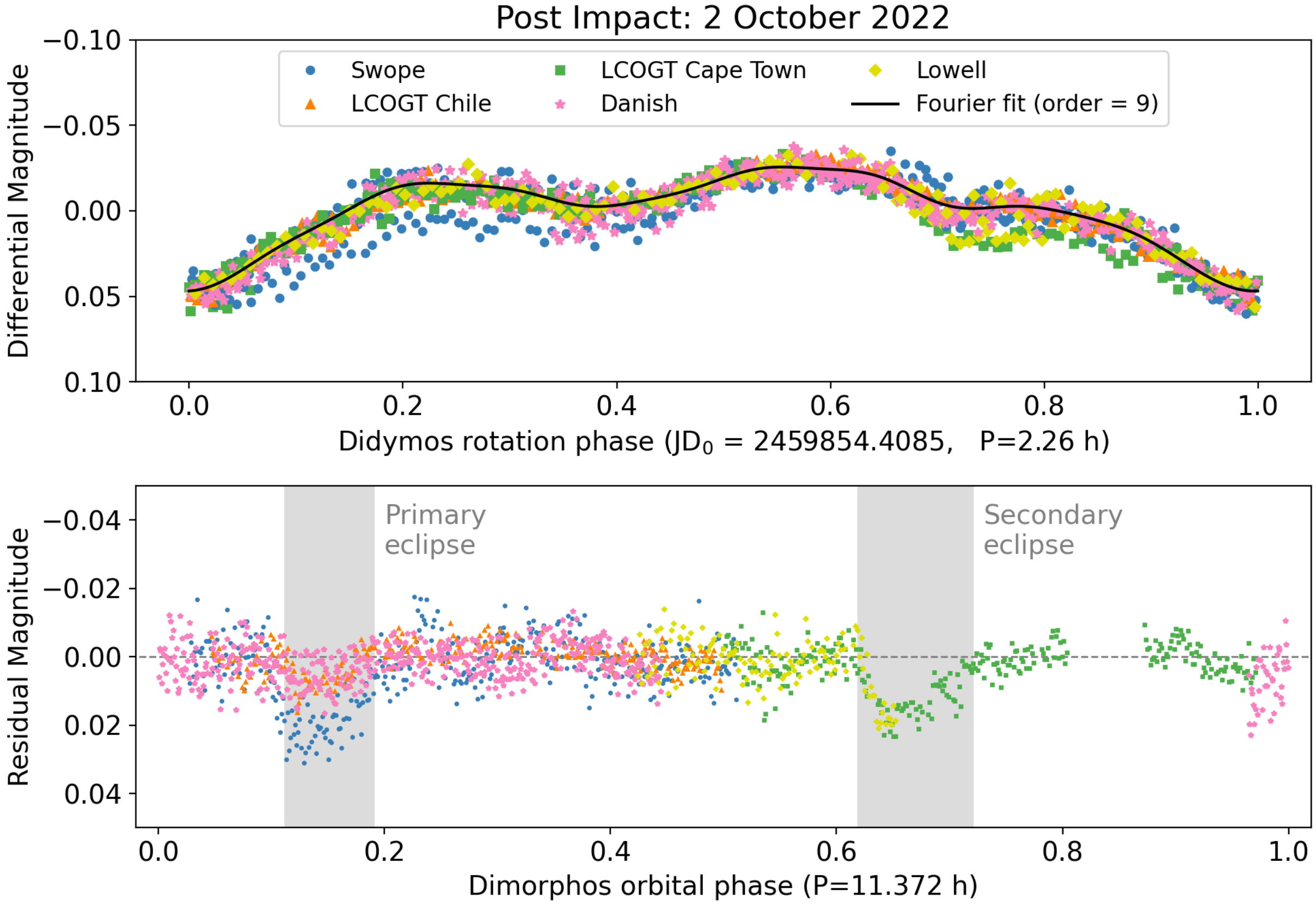

Measured photometry from UTC 2 October 2022 phase folded to the 2.26 h rotation period of Didymos (top), and the extracted mutual events (= observed data - 9th order Fourier fit to Didymos' rotation) phase folded to the new orbit period of Dimorphos (bottom). These lightcurves, collected from five different telescopes, show photometric accuracy similar to all the lightcurve data sets in our analysis. The mutual event times are highly consistent across these data sets, though residual systematics in the photometry result in slightly different event depths.

Figure 4:

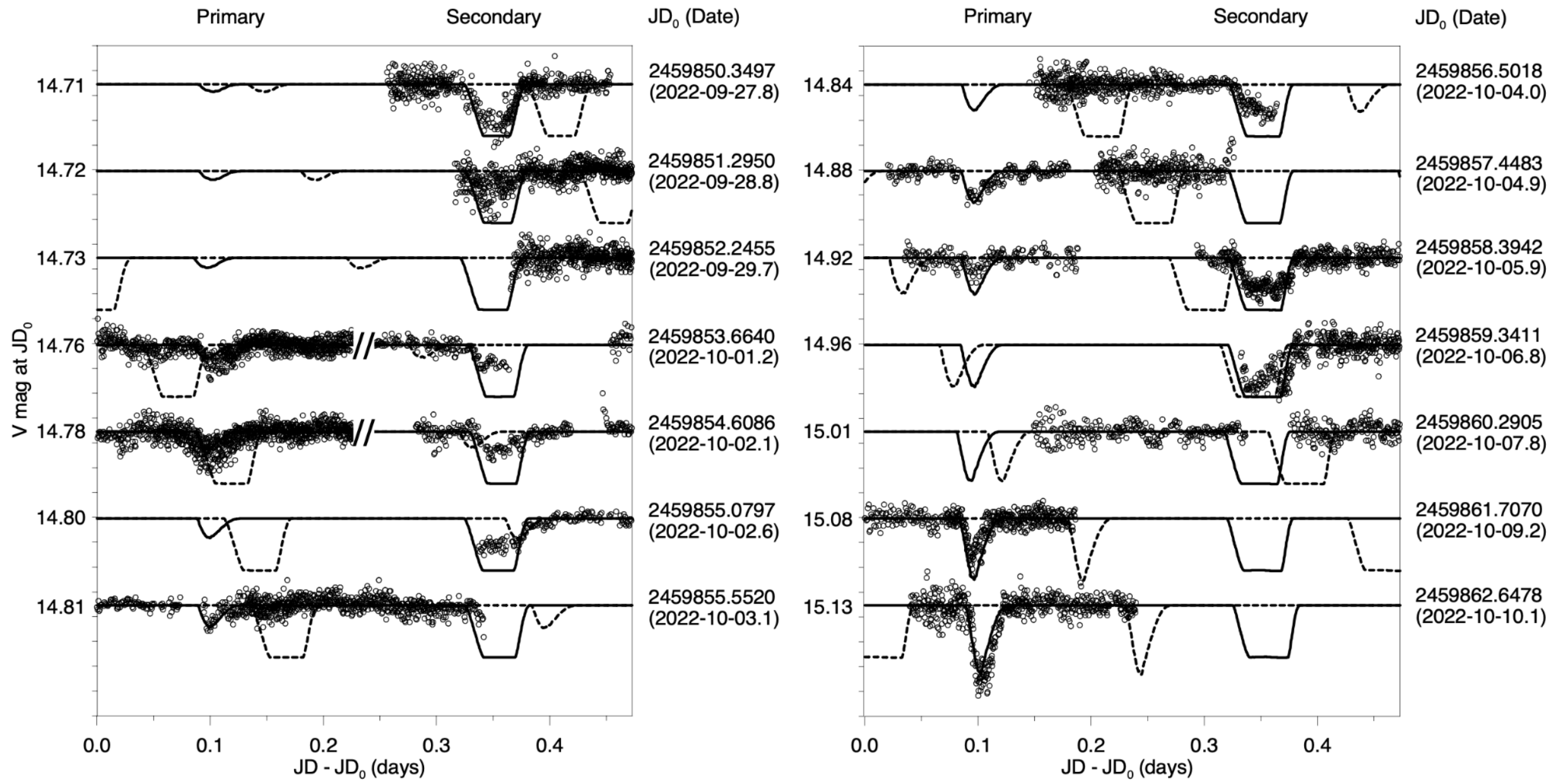

The two panels show the observed mutual events of the Didymos system. The data are marked as circles and the solid curve represents the synthetic lightcurve for the best-fit post-impact solution. The dashed curve is the pre-impact orbit prediction from ref. [6]. The primary and secondary events are shown on the left and right sides of the plots, respectively. In some cases, the observations of a secondary event precede those of a primary event (*i.e.*, their order in the data set is the inverse of that shown in the plot). We present these events in reverse order and they are separated by a "//" symbol in the plot (0.4728 day is to be subtracted from the x-coordinate of data points to the right from this separator). The y-axis shows the magnitude on the night of the observation for each data set and each tick mark has a range of 0.02 magnitudes.

# Methods

The models incorporated three types of observations of the Didymos-Dimorphos system: photometric lightcurves, radar, and Didymos-relative astrometry from DART's DRACO camera[20]. We determined the post-impact orbital period using two separate models (ref. [6], hereafter SP22, and [5,11]; hereafter N22+). Both approaches use the same sets of pre and post-impact lightcurves (Extended Data Tables 1 & 2). The SP22 approach models the lightcurve observations to determine the properties of the orbit. The N22+ approach incorporates Didymos-relative astrometry from DRACO optical navigation images to update the orbital parameters of the pre-impact orbit and includes lightcurve mutual event timings and radar observations for the post-impact solution (Extended Data Tables 3-7).

## Photometric Lightcurve Data & Reductions

Previous observations of the Didymos system[4] demonstrated the need for requirements on the photometry used in the analysis. We define our data quality requirement as an RMS < 0.01 magnitudes, where the RMS value refers to the consistency over the nightly run and results in a minimum signal-to-noise (SNR) on the individual exposures of ~100. For an accurate decomposition of the lightcurve, we require adequate coverage of the primary lightcurve outside of mutual events. We prefer two complete rotation periods of the primary ($P_{rot}$=2.26 hr) outside of the events and estimate this requirement as 6 hours of continuous observation. The observations can be split between multiple stations. Four observatories contributed data that met the photometric requirements to the lightcurve dataset for the orbital period change (Extended Data Table 2): Las Campanas Observatory 1-m Swope Telescope, the Las Cumbres Observatory global telescope network 1-m telescopes, the Danish 1.54-m telescope at the European Southern Observatory's La Silla site, and the Lowell Observatory 1.1-m Hall telescope.

The Las Campanas Observatory Swope 1-m telescope is located in the Atacama Desert, Chile[22] . The Swope 4K CCD is a visible-wavelength, direct-imaging CCD with a 29.7 x 29.8 arc-minute field of view. Swope observations were taken in the Sloan-r' filter and used sidereal tracking with 1 or 2 sky pointings each night. Instrumental aperture photometry was performed on every frame using the python package SEP[23]. We use the astroquery Python package to query Vizier[24] and Horizon[25] databases to identify *Gaia* stars and to obtain the coordinates of the asteroid for the given date of the images, respectively, and the gaiaxpy Python package, to request and download synthetic photometry of Gaia stars[26] in Sloan-r band when available. The Swope data show discrepancies in the photometry (as seen in Figure 2) at the ~0.01-0.02 mag level. There are no issues on the timing of the events, which are the key drivers for the derivation of the new orbit period. Additional reductions of this data with optimized apertures will be used to address these discrepancies.

The Las Cumbres Observatory global telescope (LCOGT) network[27] consists of telescopes at seven sites around the world, operated robotically using dynamical scheduling software[28]. We used the 1-m telescopes at the South Africa and Chile nodes with the telescopes tracking at half of the ephemeris rates. These observations were scheduled and reduced using the NEOexchange Target and Observation Manager and data reduction pipeline[29]. Images were pre-processed using the Python-based BANZAI pipeline[30]. Astrometry and photometry was performed using the Python-based NEOexchange pipeline[29]. The LCOGT data was primarily obtained in PanSTARRS-*w* band (equivalent to a broad $g + r + i$ band) and was calibrated to the *Gaia*-DR2[31] using calviacat[32], with the *w* band treated as an *r* band. Calibration stars were constrained to have "solar-like" colors.

The Danish 1.54-m telescope is located at the European Southern Observatory's La Silla site in Chile. Observations were performed by the MiNDSTEp (Microlensing Network for the Detection of Small Terrestrial Exoplanets) consortium. The Danish Faint Object Spectrograph and Camera (DFOSC) instrument, with field of view 13.7' x 13.7', was used in imaging mode. Images were taken with the Bessell *R* filter using sidereal tracking. Data reduction used a custom Python pipeline, including alignment of frames using Astrometry.net tools[33]. Relative photometry was calibrated using the procedure outlined in ref. [34] using the calviacat[32] package and the *Gaia* DR3 star catalog, with conversion to SDSS-*r* band magnitudes assuming a color of $(g\text{-}r)=0.52$ for Didymos[4,35].

The Lowell 1.1-m Hall telescope, located on Anderson Mesa south of Flagstaff, Arizona, is equipped with a 4k x 4k CCD that images a 25 arcmin square field. The telescope was tracked at half of the ephemeris rate. Exposures were taken with a broad *VR*-band filter. Photometric calibration was based on field star magnitudes from the PanSTARRS catalog. Only stars with high signal-to-noise (>100) and solar-like colors were used for calibration. For the 2022-10-02 data, the photometry was measured using the Canopus software package. For the 2022-10-05 data, the photometry was measured using the PhotometryPipeline[36].

We added lightcurve observations from three telescopes (Table 1) to augment the pre-impact lightcurve solutions published in ref. [6] and [5]: the 6.5-m Magellan Baade telescope, the SOAR (Southern Astrophysical Research) 4.1-m telescope, and the 4.3-m Lowell Discovery Telescope. Both of the updated models confirmed the previous solutions.

## Lightcurve Decomposition

To model the photometric data of the binary asteroid system, we follow the decomposition methods defined in ref. [17,37] and discussed in ref. [4]. Outside of mutual events, the largest signal in the Didymos system lightcurve is the flux of the primary which can be represented by the following Fourier series:

$$F_1(t) = C_1 + \sum_{k=1}^{m_1} \left[ C_{1k} \cos \frac{2\pi k}{P_1} (t - t_0) + S_{1k} \sin \frac{2\pi k}{P_1} (t - t_0) \right]$$

$F_1(t)$ is the flux of the primary, Didymos, at time t, $C_1$ is the mean flux of the primary, $C_{1k}$ and $S_{1k}$ are the Fourier coefficients, $P_1$ is the lightcurve rotational period of Didymos, $t_0$ is the zero-point time, and $m_1$ is the maximum significant order. By using this mathematical representation for the system, we assume that Didymos is in principal axis rotation, that mutual illumination between the objects is negligible, and that the rotational lightcurve does not change with time.

The lightcurve data is corrected to constant geocentric and heliocentric distances and a consistent solar phase angle. We connect data from different telescopes by scaling them in relative magnitude compared to each other, which has no impact on the timing of the mutual events.

We use observations taken outside of mutual events to fit the rotational lightcurve of Didymos. The rapidly changing Earth-Didymos-Sun geometry during this period of Didymos' close approach to Earth causes observable changes in the primary rotational lightcurve. For our previous work[4], we were able to combine data on the timescales of days to weeks. For this dataset, separate decompositions are done for each Julian Day (JD). We correct for the overall fading of the ejecta for each dataset by fitting a linear flux trend before performing the lightcurve decomposition.

## Radar Observations

We observed Didymos and Dimorphos using the Goldstone X-band radar (3.5 cm, 8560 MHz) on the 70-m DSS-14 telescope on 14 dates between 27 September - 13 October 2022. On 2, 6, and 9 October, we also used the 100-m Green Bank Telescope to receive radar echoes in a bistatic configuration with transmissions from Goldstone. Typical transmitter power was 430 kW. We obtained echo power spectra during each of the observing windows and range-Doppler images on several days centered on 4 October when the signal to noise ratios (SNRs) were the highest. Didymos was clearly detected in all of the data ($> 3\sigma$) and its maximum bandwidth varied from 22 Hz on 27 September, when the subradar latitude was -50 degrees, to 34 Hz on 13 October, when its subradar latitude was -32 degrees (based on the pole direction estimated by ref. [11]).

Detecting Dimorphos was challenging and required experimenting with setups having different frequency resolutions, range resolutions, and integration times. This process was a trade-off between obtaining longer integrations with sufficiently high SNRs to detect Dimorphos versus reducing the smearing caused by the orbital motion during the integration. We found that the echo from Dimorphos was most consistently visible at resolutions of 1 Hz in the echo power spectra and at 0.5 Hz in the images. Due to the 11.9 h rotation period, a diameter of ~160 meters, and a subradar latitude of -50 to -30 deg[11], the echo from Dimorphos was expected to have a bandwidth of about 1 Hz[11], so the data do not resolve Dimorphos in frequency but maximize the SNRs by nearly matching the bandwidth. The contribution of self-noise in the echo power spectra is negligible and does not affect the SNRs significantly. We attempted imaging with time delay resolutions of 0.5 μs and 1 μs (corresponding to range resolutions of 75 m and 150 m), and found that the 0.5 μs setup yielded more consistent detections. We experimented with summing data spanning a range of time intervals and found that the echo from Dimorphos was not clearly visible in all the data on any given day. It became more difficult to detect Dimorphos after 4 October as the distance to Didymos increased and the SNRs correspondingly decreased. Figure 2 shows range-Doppler images and Extended Data Figure 5 shows selected echo power spectra in which the echo from Dimorphos was seen.

We measured the separations between Dimorphos and Didymos in the echo power spectra and range-Doppler images and used these measurements in the estimation of the orbital parameters of Dimorphos relative to Didymos. The separations in Doppler frequency and range between Didymos and Dimorphos relate to the relative velocity and distance along the observer's line of sight due to their mutual orbit about each other. We used only data in which both Didymos and Dimorphos were clearly visible for making these measurements. The echo power spectra were processed so that hypothetical echoes from the Didymos system barycenter appear at 0 Hz[38]. Because the reflex motion of Didymos about the system barycenter is < 10 m (0.08 Hz)[11], we assumed that the Didymos center of mass (COM) is at 0 Hz so that the Doppler frequency of Dimorphos represents the relative Doppler shift. The echo from Dimorphos is unresolved so we assumed that its COM was located in the Doppler bin that contained the strongest spike due to the echo from Dimorphos. We assigned uncertainties of -± 2 Hz to the Doppler separation measurements to take into account the uncertainties due to the frequency resolution of the spectra (1 Hz), the ephemeris errors in the location of the system barycenter (0.24 Hz, 3$\sigma$), and the reflex motion of Didymos about the system barycenter (< 0.1 Hz). Consequently, the principal source of uncertainty in measurements of the range-Doppler separations are the Doppler frequencies of Dimorphos.

Due to the low SNRs, the COM of Didymos is hard to locate in the range-Doppler images, so we assumed it is located 375 m (5 range pixels at 75 m/pixel) behind the leading edge, which is the brightest part of the echo and easiest to see. This distance equals the equatorial radius reported from the 3D shape model obtained by ref. [11] and is consistent with preliminary estimates from the DART spacecraft images reported by ref. [16]. The echo from Dimorphos extended over one to three range rows and we assumed that its COM is in the trailing row. We assigned uncertainties of 150 m (two range rows) to the range separation measurements. Tables 5 and 6 show the range and Doppler frequency of Dimorphos relative to Didymos that were used in the orbit determination. We estimated 8 range measurements on 9 October (when reception at Green Bank facilitated detecting echoes from Dimorphos), far more than on any other day, so we inflated their uncertainties by a factor of 3 in order to mitigate the effects of correlated errors.

## Didymos-Relative Optical Astrometry from DRACO images

We measured the positions of Dimorphos relative to Didymos in 16 DRACO images taken in the minutes prior to impact on 26 September 2022 between 23:10:58.235 and 23:12:39.336 UTC to use in the orbit estimation process. At the time these measurements were made, no shape models estimated from spacecraft images were available to fit to the partially illuminated figures of the two bodies, so we measured the intersections of the limbs with the relative position vectors. These measurements were differenced to estimate the limb-to-limb positions of Dimorphos relative to Didymos. These positions were mapped from image coordinates into Right Ascension (RA) and Declination (DEC) using the camera model and the GNC (Guidance, Navigation, and Control) spacecraft attitude knowledge. Measurement uncertainties of 1.13 x $10^{-3}$ degrees (3$\sigma$) were derived by repeating this process and comparing the different observations. We assumed the equatorial extents of Didymos and Dimorphos to be 425 m and 88 m respectively and added an angular distance corresponding to 425 – 88 = 337 m (±20 m 1$\sigma$) uncertainty) in the direction of the limb-to-limb separations to estimate the distances between the COMs. Since the measurements covered a very short time span, we de-weighted the uncertainties by 4x (√16) to mitigate effects of correlated measurement errors. We de-weighted the DEC measurements by an additional factor of two because they are clearly noisier than the RA measurements. Extended Data Table 7 lists the observations and uncertainties.

**Orbital Period Determination Via Lightcurves (SP22 Method)**

The ref. [6] numerical model of the Didymos system was developed using the techniques described in ref. [39–41]. Didymos and Dimorphos are represented by ellipsoids with axial ratios of $a_1/c_1=b_1/c_1=1.37$, $a_2/c_2=1.53$, $b_2/c_2=1.50$[16]. The motion of the two bodies is assumed to be Keplerian. The post-impact system was analyzed with no *a priori* assumption on the new binary orbital period. The lightcurve data from 28 & 29 September showed that parts of the data were attenuated with respect to the primary's rotational lightcurve. Those sections of the data were iteratively masked until all of the data points in the mutual events were identified and the lightcurve decomposition was complete. The first mutual event (0.03 magnitudes deep) was determined to be a secondary eclipse since the system geometry predicted very shallow or absent primary events.

We adapted the method from ref. [6] to estimate the uncertainty of the post-impact period. When stepping the period over a suitable interval we computed normalized $\chi^2$ for each step. We determined its 3-σ uncertainty as an interval in which $\chi^2$ is below a certain limit. The adopted limiting p-value corresponds to the probability that the $\chi^2$ exceeds a particular value only by chance equal to 0.27%. At each step of the period scanning, the mean anomaly of Dimorphos at the epoch of the impact was also scanned within its 3-σ uncertainty interval that was determined by ref. [6] and that we have updated using the additional data taken in July 2022. The SP22 pre-impact period was 11.921478 ± 0.000123 (3σ) hours.

The SP22 model determines a post-impact period of 11.372 ± 0.017 (3σ) hours corresponding to an orbit period change of -33.0 ± 1.0 (3σ) minutes.

**Orbital Period Determination via Radar and Lightcurves (N22+ Method)**

The lightcurve analysis method described in ref. [5] is a less complicated approach compared to the methods presented in ref. [6]. However, it has the advantage of combining information from different data types such as radar, relative optical astrometry from DRACO images, and lightcurve mutual events. The pre-impact orbital period using the N22+ approach was 11.92148 ± 0.00013 (3σ) hours.

Lightcurve decomposition was done independently from the SP22 process and required identifying mutual events. The first identified post-impact mutual event was on UTC 28 September 2022. We expected that the head-on impact would decrease the orbital period compared to the pre-impact solution and expected an event with a length of approximately 1 hour. To identify the mutual event, we tested a range of orbit periods from 11-12 hours in time steps of 0.1 hours with a best match of 11.4 hours. Subsequent observations helped refine the initial estimate.

For each mutual event there are four contact times: when the event begins and flux decreases ($T_1$), when flux reaches a minimum ($T_2$), when the flux begins to increase ($T_3$), and when the event ends and the flux returns to the baseline ($T_4$). We use times $T_{1.5}$ and $T_{3.5}$ in the orbit determination. These times are when the flux is at half the total drop in flux during the event (Fig 1 in ref. [5]). We use 1$\boldsymbol{\sigma}$ uncertainties of ($T_{1.5}$ - $T_1$)/2 and ($T_4$ - $T_{3.5}$)/2 for $T_{1.5}$ and $T_{3.5}$, respectively.

We used a least-squares approach, as described in ref. [5], for estimating the orbital parameters of Dimorphos relative to Didymos. Prior to the DART impact, Dimorphos is assumed to be a point mass on a modified Keplerian orbit around Didymos with an additional term for modeling the drift in mean motion due to nongravitational effects such as the Binary YORP effect and tidal dissipation. The post-impact orbit was assumed to be Keplerian, since the data-arc length is too short to detect a drift in mean motion. We used $\Delta n$ to capture the change in mean motion due to the DART impact. The mean anomaly, $M$, and mean motion, $n$, of Dimorphos at time, $t$, are given by:

$$M(t) = M_0 + n_0(t - t_0) + \frac{1}{2}\dot{n}(t - t_0)^2 \text{ for } t < t_{imp}$$

$$M(t) = M_{imp} + (n_{imp} + \Delta n)(t - t_{imp}) \text{ for } t > t_{imp}$$

$$n(t) = n_0 + \dot{n}(t - t_0) \text{ for } t < t_{imp}$$

$$n(t) = n_{imp} + \Delta n \text{ for } t > t_{imp}$$

Where $t_{imp}$ is time of the DART impact, $M_0$ and $n_0$ are the mean anomaly and mean motion at $t_0$, $\dot{n}$ is the linear drift in mean motion due to nongravitational effects, and $M_{imp}$ and $n_{imp}$ are the mean anomaly and mean motion at impact.

We used differential corrections as described in ref. [5] for estimating the orbital parameters $M_0$, $n_0$, $\dot{n}$, $\Delta n$, the pre-impact semimajor axis ($a$), and the orbit pole longitude ($\lambda$) and latitude ($\beta$). This requires calculating a computed value corresponding to each observation using a model. We used three kinds of observations: lightcurve mutual event times, radar range and Doppler measurements of Dimorphos relative to Didymos, and the separation of Dimorphos from Didymos as seen in spatially-resolved DRACO images. The modeling of the first two observables is described in ref. [5]. In order to model the separation of Dimorphos from Didymos in DRACO images, we used SPICE[42] to subtract the RA and DEC of the COM of Didymos from those of the COM of Dimorphos as seen from the DART spacecraft.

The N22+ approach results in a post-impact period of 11.371 ± 0.016 (3σ) hours and an orbit period change of -33.0 ± 1.0 (3σ) minutes. The best fit orbit parameters are presented in Table 3.

## Methods References

## Data Availability

The lightcurves and radar data used in this analysis of the orbital period are available in the JHU/APL Data archive at: https://lib.jhuapl.edu/papers/orbital-period-change-of-dimorphos-due-to-the-dart/. The DRACO images can be found in an archive associated with the Daly et al. paper (https://lib.jhuapl.edu/papers/dart-an-autonomous-kinetic-impact-into-a-near-eart/).

In addition, all observations from Las Campanas Observatory, Las Cumbres Observatory global telescope (LCOGT) network, and the Lowell Discovery Telescope will be publicly archived at the Planetary Data System Small Bodies Node with the DART mission data by October 2023. The radar datasets will be separately archived at the Planetary Data System.

## Code Availability

The algorithms used here were published in Scheirich & Pravec (2022) and Naidu et al. (2022).

## Acknowledgements

This work was supported by the DART mission, NASA Contract No. 80MSFC20D0004. Part of this research was carried out at the Jet Propulsion Laboratory, California Institute of Technology, under a contract with the National Aeronautics and Space Administration. The work by P.S. and P.P. was supported by the Grant Agency of the Czech Republic, grant 20-04431S. They appreciate access to computing and storage facilities owned by parties and projects contributing to the National Grid Infrastructure MetaCentrum provided under the program "Projects of Large Research, Development, and Innovations Infrastructures" (CESNET LM2015042) and the CERIT Scientific Cloud LM2015085. The Green Bank Observatory is a facility of the National Science Foundation operated under cooperative agreement by Associated Universities, Inc. Observations at the Danish 1.54m telescope were supported, in part, by the European Union H2020-SPACE-2018-2020 research and innovation programme under grant agreement No. 870403 (NEOROCKS). This work makes use of observations from the Las Cumbres Observatory global telescope network. This paper includes data gathered with the 6.5 meter Magellan Telescopes located at Las Campanas Observatory, Chile. Based on observations obtained at the Southern Astrophysical Research (SOAR) telescope, which is a joint project of the Ministério da Ciência, Tecnologia e Inovações do Brasil (MCTI/LNA), the US National Science Foundation's NOIRLab, the University of North Carolina at Chapel Hill (UNC), and Michigan State University (MSU). These results made use of the Lowell Discovery Telescope (LDT) at Lowell Observatory. Lowell is a private, non-profit institution dedicated to astrophysical research and public appreciation of astronomy and operates the LDT in partnership with Boston University, the University of Maryland, the University of Toledo, Northern Arizona

University and Yale University. U.G.J. acknowledges funding from the Novo Nordisk Foundation Interdisciplinary Synergy Programme grant no. NNF19OC0057374, and from the European Union H2020-MSCA-ITN-2019 grant no. 860470 (CHAMELEON). E.K. is supported by the National Research Foundation of Korea 2021M3F7A1082056. P.L.P. was partly funded by "Programa de Iniciación en Investigación-Universidad de Antofagasta. INI-17-03".

## Author Contributions

C.A.T. is the lead of the DART mission's Observations Working Group. She coordinated observations, led the paper writing, and participated in the observing. S.P.N and P.S performed the independent modeling efforts to determine the post-impact period and period change. N.A.M. and P.P. accepted lightcurve data based on the requirements and performed the decompositions. S.R.C. supported orbit estimation of Dimorphos. A.S.R., N.L.C., and A.F.C. lead the DART Investigation Team, contributed to the writing and revision of this paper, and coordinated inputs across the DART Investigation Team. D.J.O., C.C., N.M. planned, executed, and reduced the data from Las Campanas Observatory's Swope telescope. T.A.L. led the data collection from the Las Cumbres Observatory global telescope network. L.A.M.B. and M.B. assisted with proposal writing, observing, and data processing for the radar observations. A.R., P.K, and K.H. performed data reduction and photometry for the Danish telescope data set. D.M. and B.R. measured the positions of in the OPNAV images. P.A.T., A.D.S., P.S., W.P.A., and G.W. helped with the Green Bank Observatory observations. M.W.B. assisted with the proposal for Green Bank Observatory. C.S., U.G.J., and M.D. planned and coordinated observations at the Danish telescope. B.S. and T.P. performed the observations and reductions for the Lowell Hall telescope. M.M.K, T.L.F., and C.E.H. performed the observations for the Lowell Discovery Telescope. J.D.G. and M.B. assisted with the planning of the radar observations. J.B. generated the spacecraft SPK that was used in the OPNAV treatment. J.C., E.G., S.G., and L.P. guilt the portal and pipeline for scheduling and reducing the LCOGT data. M.B., M.J.B., E.K., P.L.-P., M.R., and S.S. performed observations at the Danish Telescope. W.H.R. and E.V.R assisted with planning the observing effort. H.F.A. provided comments on the manuscript and performed the formatting.

## Competing Interests

The authors declare no competing interests.

## Additional Information

Correspondence should be addressed to Dr. Cristina Thomas (cristina.thomas@nau.edu)

# Extended Data

Extended Data Figure 1:

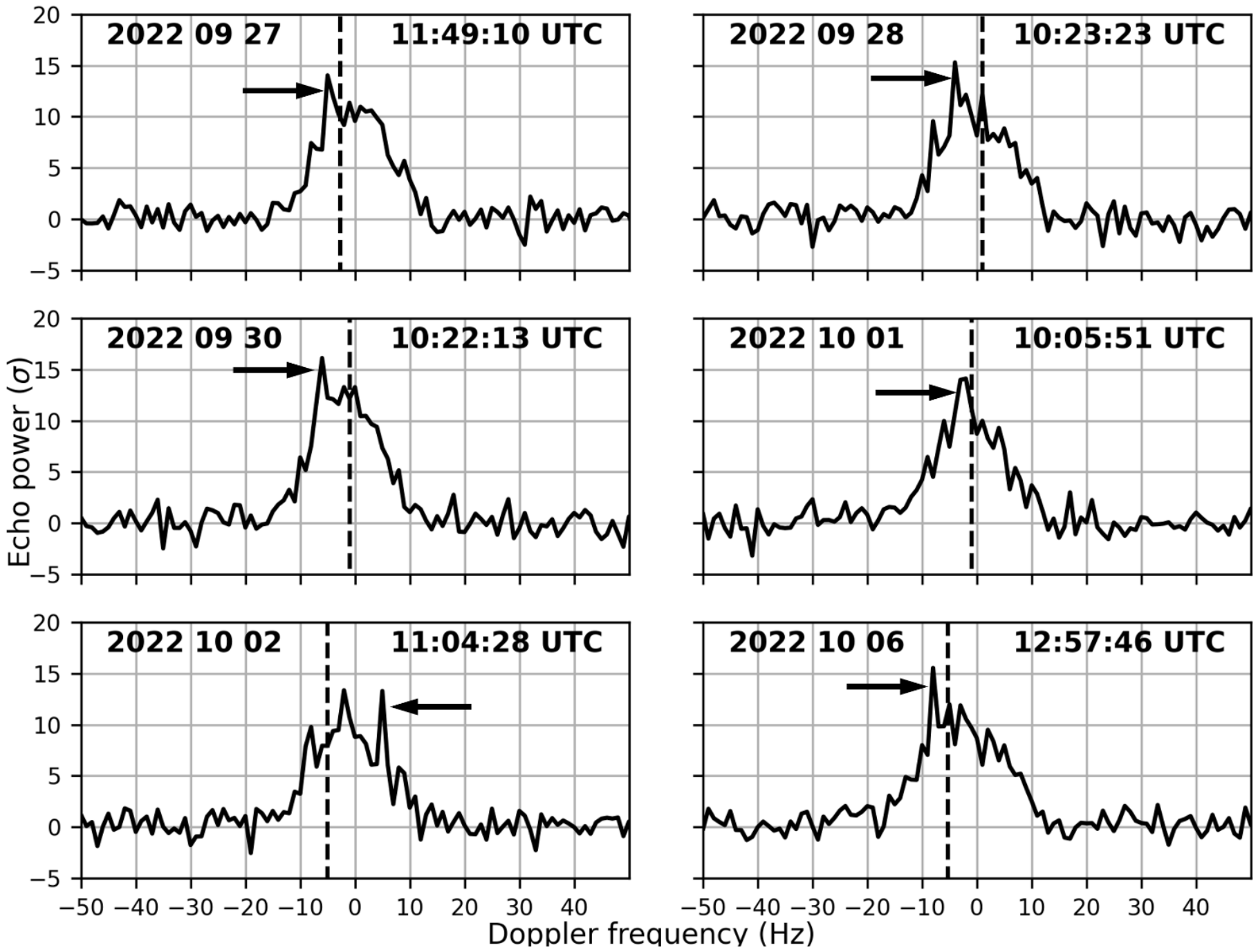

Selected radar echo power spectra obtained at Goldstone that were used to measure the Doppler separations in Table 6. The spectra were obtained in the opposite sense of circular polarization as the transmitted wave. Each spectrum was integrated for 10-15 minutes in order to detect Dimorphos with minimum smear due to orbital motion (< 8 degrees). Echoes from Didymos are centered on 0 Hz and have a bandwidth of between 22-34 Hz. The echo from Dimorphos appears as a narrow spike superimposed on the signal from Didymos, a pattern observed with radar observations of dozens of other near-Earth asteroids (e.g.,[43]), indicated by the arrows. The Doppler frequency of Dimorphos varies with time between positive and negative values due to its orbital motion and estimated values can be found in Table 6. Dashed vertical lines show the Doppler frequencies of Dimorphos predicted by the pre-impact orbit. Prediction uncertainties are smaller than the resolution of the spectra.

**Extended Data Table 1:**

Additional pre-impact photometric observations of (65803) Didymos beyond those described in Pravec et al. (2022).

| Date (UT) | Start Time (UT) | Duration (Hr) | # of Points | Telescope | RMS Residual (N22+) mag | RMS Residual (SP22) mag |
|---|---|---|---|---|---|---|
| 2022-07-02 | 03:59 | 6.6 | 193 | LCO/Magellan Baade 6.5-m | 0.008 | 0.009 |
| 2022-07-04 | 06:52 | 3.8 | 129 | CTIO/SOAR 4.1-m | 0.007 | 0.006 |
| 2022-07-05 | 04:24 | 6.4 | 210 | CTIO/SOAR 4.1-m | 0.008 | 0.006 |
| 2022-07-06 | 08:02 | 3.2 | 89 | Lowell Discovery Telescope 4.3-m | 0.005 | 0.006 |
| 2022-07-07 | 07:50 | 3.5 | 85 | Lowell Discovery Telescope 4.3-m | 0.009 | 0.006 |

**Extended Data Table 2:**

Post-impact photometric observations of (65803) Didymos used to derive the new orbital period and period change as a result of impact.

| Date (UT) | Start Time (UTC) | End Time (UTC) | Duration (Hr) | # of Points | Telescope | RMS Residual (N22+) mag | Slope Correction (N22+) mag/day | RMS Residual (SP22) mag | Slope Correction (SP22) mag/day |
|---|---|---|---|---|---|---|---|---|---|
| 2022-09-28 | 2:33 | 6:09 | 3.6 | 237 | LCO/Swope 1-m | 0.008 | 0.10 | 0.008 | 0.07 |
| 2022-09-28 | 2:38 | 9:16 | 6.7 | 340 | La Silla/Danish 1.54-m | 0.006 | 0.10 | 0.006 | 0.12 |
| 2022-09-29 | 2:40 | 9:17 | 6.6 | 433 | LCO/Swope 1-m | 0.007 | 0.24 | 0.007 | 0.25 |
| 2022-09-29 | 2:50 | 9:33 | 6.7 | 639 | La Silla/Danish 1.54-m | 0.005 | 0.10 | 0.005 | 0.12 |
| 2022-09-29 | 4:52 | 8:47 | 3.9 | 212 | CTIO/LCOGT-LSC 1-m | 0.003 | 0.12 | 0.003 | 0.12 |
| 2022-09-30 | 2:40 | 9:41 | 7.0 | 669 | La Silla/Danish 1.54-m | 0.006 | 0.10 | 0.006 | 0.11 |
| 2022-09-30 | 2:52 | 9:15 | 6.4 | 420 | LCO/Swope 1-m | 0.008 | 0.20 | 0.008 | 0.20 |
| 2022-09-30 | 3:52 | 9:16 | 5.4 | 319 | CTIO/LCOGT-LSC 1-m | 0.004 | 0.12 | 0.004 | 0.12 |
| 2022-09-30 | 21:45 | 1:12 | 3.5 | 168 | SAAO/LCOGT-CPT 1-m | 0.004 | 0.22 | 0.004 | 0.21 |
| 2022-10-01 | 3:28 | 9:14 | 5.8 | 376 | LCO/Swope 1-m | 0.007 | 0.22 | 0.006 | 0.23 |
| 2022-10-01 | 4:00 | 9:11 | 5.2 | 292 | CTIO/LCOGT-LSC 1-m | 0.005 | 0.06 | 0.004 | 0.08 |
| 2022-10-01 | 6:33 | 9:28 | 2.9 | 278 | La Silla/Danish 1.54-m | 0.005 | 0.25 | 0.005 | 0.10 |
| 2022-10-01 | 21:48 | 3:09 | 5.4 | 268 | SAAO/LCOGT-CPT 1-m | 0.005 | 0.08 | 0.005 | 0.06 |
| 2022-10-02 | 3:05 | 8:33 | 5.5 | 530 | La Silla/Danish 1.54-m | 0.006 | 0.00 | 0.005 | 0.09 |
| 2022-10-02 | 3:15 | 9:19 | 6.1 | 359 | LCO/Swope 1-m | 0.007 | 0.15 | 0.006 | 0.16 |
| 2022-10-02 | 4:00 | 9:11 | 5.2 | 269 | CTIO/LCOGT-LSC 1-m | 0.004 | 0.02 | 0.003 | 0.00 |
| 2022-10-02 | 8:19 | 10:55 | 2.6 | 132 | Lowell/Hall 1.1-m | 0.005 | 0.15 | 0.006 | 0.09 |
| 2022-10-02 | 22:00 | 00:56 | 2.9 | 136 | SAAO/LCOGT-CPT 1-m | 0.003 | 0.00 | 0.003 | 0.00 |
| 2022-10-03 | 1:00 | 3:02 | 2.0 | 99 | SAAO/LCOGT-CPT 1-m | 0.003 | 0.10 | 0.003 | 0.13 |
| 2022-10-03 | 3:29 | 9:27 | 6.0 | 385 | LCO/Swope 1-m | 0.007 | 0.12 | 0.007 | 0.12 |
| 2022-10-03 | 4:05 | 8:42 | 4.6 | 248 | CTIO/LCOGT-LSC 1-m | 0.004 | 0.12 | 0.004 | 0.10 |
| 2022-10-03 | 4:27 | 5:28 | 1.0 | 98 | La Silla/Danish 1.54-m | 0.006 | 0.20 | 0.005 | 0.12 |
| 2022-10-04 | 3:46 | 8:24 | 4.6 | 224 | LCO/Swope 1-m | 0.006 | 0.15 | 0.006 | 0.11 |
| 2022-10-04 | 4:15 | 8:47 | 4.5 | 248 | CTIO/LCOGT-LSC 1-m | 0.005 | 0.05 | 0.005 | 0.04 |
| 2022-10-04 | 23:15 | 3:04 | 3.8 | 206 | SAAO/LCOGT-CPT 1-m | 0.005 | 0.00 | 0.004 | 0.00 |
| 2022-10-05 | 3:40 | 5:51 | 2.2 | 151 | La Silla/Danish 1.54-m | 0.006 | 0.10 | 0.006 | 0.00 |
| 2022-10-05 | 3:45 | 6:39 | 2.9 | 181 | LCO/Swope 1-m | 0.009 | 0.25 | 0.008 | 0.28 |
| 2022-10-05 | 22:18 | 1:59 | 3.7 | 194 | SAAO/LCOGT-CPT 1-m | 0.006 | 0.04 | 0.005 | 0.00 |
| 2022-10-06 | 3:49 | 9:20 | 5.5 | 346 | LCO/Swope 1-m | 0.010 | 0.06 | 0.009 | 0.06 |
| 2022-10-06 | 4:31 | 8:09 | 3.6 | 194 | CTIO/LCOGT-LSC 1-m | 0.005 | 0.10 | 0.004 | 0.10 |
| 2022-10-06 | 5:03 | 9:27 | 4.4 | 370 | La Silla/Danish 1.54-m | 0.005 | 0.00 | 0.005 | 0.14 |
| 2022-10-07 | 4:10 | 8:28 | 4.3 | 379 | La Silla/Danish 1.54-m | 0.006 | 0.12 | 0.006 | 0.11 |
| 2022-10-07 | 4:52 | 9:08 | 4.3 | 219 | CTIO/LCOGT-LSC 1-m | 0.007 | 0.10 | 0.007 | 0.09 |
| 2022-10-07 | 22:31 | 3:03 | 4.5 | 246 | SAAO/LCOGT-CPT 1-m | 0.007 | 0.05 | 0.006 | 0.05 |
| 2022-10-08 | 4:00 | 9:29 | 5.5 | 471 | La Silla/Danish 1.54-m | 0.007 | 0.15 | 0.006 | 0.06 |
| 2022-10-08 | 22:37 | 3:02 | 4.4 | 200 | SAAO/LCOGT-CPT 1-m | 0.010 | 0.05 | 0.011 | 0.00 |
| 2022-10-09 | 4:49 | 9:12 | 4.4 | 244 | CTIO/LCOGT-LSC 1-m | 0.006 | 0.00 | 0.005 | 0.00 |
| 2022-10-09 | 6:41 | 9:26 | 2.8 | 244 | La Silla/Danish 1.54-m | 0.007 | 0.00 | 0.005 | 0.06 |
| 2022-10-10 | 4:30 | 9:18 | 4.8 | 419 | La Silla/Danish 1.54-m | 0.006 | 0.12 | 0.006 | 0.06 |

Extended Data Table 3: Best-fit orbit parameters using the N22 method. The input observations data is listed in Tables 4, 5, 6, and 7.

| **Parameter** | **Estimate +/- 1$\sigma$ uncertainties** |
| --- | --- |
| Epoch (UTC) | 2022 Sep 26 23:14:24.183 |
| Orbit pole longitude ($\lambda$, degrees) | 313.3 +/- 5.2 |
| Orbit pole latitude ($\beta$, degrees) | -79.3 +/- 1.0 |
| Pre-impact semimajor axis ($a$, km) | 1.206 +/- 0.035 |
| Mean anomaly at epoch ($M_0$, degrees) | 178.9 +/- 5.5 |
| Pre-impact period (h) | 11.921473 +/- 0.000044 |
| Mean motion at epoch ($n_0$, rad/sec) | (1.4640214 +/- 0.0000054) x $10^{-4}$ |
| Rate of change of mean motion ($\dot{n}$, rad/sec$^2$) | (5.4 +/- 1.6) x $10^{-18}$ |
| Post-impact period (h) | 11.3712 +/- 0.0055 |
| Period change (min) | -33.02 +/- 0.33 |
| Change in mean motion ($\Delta n$, rad/sec) | (7.085 +/- 0.070) x $10^{-6}$ |

Note: Formal uncertainties are scaled by 2x in order to capture errors from unmodeled sources.

Extended Data Table 4: Mutual event times measured in post-impact lightcurves for the N22+ approach. All times are one-way light-time corrected to reflect the time of the events at the asteroid, not the times that they were observed from Earth. The beginnings and ends of events correspond to $T_{1.5}$ and $T_{3.5}$. The fourth column shows the post-fit residuals (observed - computed) for the solution in Table 3, normalized by the 1-sigma uncertainty listed in the third column. The fifth column shows the time since impact.

| Time (UTC) | Event type | Unc. (days) | Residuals (sigma) | $\Delta t_{impact}$ (days) |
|---|---|---|---|---|
| 2022 SEP 28 04:28:07 | Beginning of secondary eclipse | 0.011 | 0.50 | 1.22 |
| 2022 SEP 28 05:14:03 | End of secondary eclipse | 0.008 | -0.94 | 1.25 |
| 2022 SEP 29 03:02:00 | Beginning of secondary eclipse | 0.01 | -0.04 | 2.16 |
| 2022 SEP 29 03:39:53 | End of secondary eclipse | 0.0135 | -1.45 | 2.18 |
| 2022 OCT 01 06:11:57 | Beginning of primary eclipse | 0.0075 | 0.58 | 4.29 |
| 2022 OCT 01 06:45:04 | End of primary eclipse | 0.008 | 0.71 | 4.31 |
| 2022 OCT 01 23:12:54 | Beginning of secondary eclipse | 0.0115 | 0.19 | 5.00 |
| 2022 OCT 02 00:08:03 | End of secondary eclipse | 0.0115 | -0.55 | 5.04 |
| 2022 OCT 02 04:43:58 | Beginning of primary eclipse | 0.0075 | -0.36 | 5.19 |
| 2022 OCT 02 05:28:45 | End of primary eclipse | 0.01 | 0.63 | 5.26 |
| 2022 OCT 02 22:57:47 | End of secondary eclipse | 0.0075 | -0.23 | 5.99 |
| 2022 OCT 04 08:13:37 | Beginning of secondary eclipse | 0.01 | 1.23 | 7.37 |
| 2022 OCT 05 00:40:10 | Beginning of primary eclipse | 0.0085 | -1.18 | 8.06 |
| 2022 OCT 05 01:22:30 | End of primary eclipse | 0.008 | -0.50 | 8.09 |
| 2022 OCT 05 23:39:33 | Beginning of primary eclipse | 0.0045 | 0.40 | 9.02 |
| 2022 OCT 05 23:56:15 | End of primary eclipse | 0.0055 | -1.89 | 9.03 |
| 2022 OCT 06 05:29:36 | Beginning of secondary eclipse | 0.005 | 1.23 | 9.26 |
| 2022 OCT 06 06:32:15 | End of secondary eclipse | 0.0055 | 0.17 | 9.30 |
| 2022 OCT 07 04:55:12 | End of secondary eclipse | 0.01 | -1.30 | 10.24 |
| 2022 OCT 08 08:34:39 | Beginning of primary eclipse | 0.0065 | 1.28 | 11.39 |
| 2022 OCT 08 08:51:21 | End of primary eclipse | 0.007 | -0.71 | 11.40 |
| 2022 OCT 09 07:12:25 | Beginning of primary eclipse | 0.0065 | 0.80 | 12.33 |
| 2022 OCT 09 07:33:35 | End of primary eclipse | 0.0085 | -0.64 | 12.35 |
| 2022 OCT 10 05:54:48 | Beginning of primary eclipse | 0.005 | 1.06 | 13.28 |
| 2022 OCT 10 06:14:58 | End of primary eclipse | 0.0055 | -1.19 | 13.29 |

Extended Data Table 5: Goldstone radar range measurements of Dimorphos relative to Didymos. The fourth column shows the post-fit residuals (observed - computed) for the solution in Table 3, normalized by the 1-sigma uncertainty listed in the third column.

| **Receive time (UTC)** | **Range (m)** | **Unc. (m)** | **Residuals (sigma)** |
|---|---|---|---|
| 2022 OCT 04 11:32:00 | -825 | 150 | -0.17 |
| 2022 OCT 04 11:55:00 | -900 | 150 | -0.36 |
| 2022 OCT 09 10:28:09 | 828 | 450 | -0.14 |
| 2022 OCT 09 10:38:09 | 965 | 450 | 0.10 |
| 2022 OCT 09 10:48:09 | 942 | 450 | 0.00 |
| 2022 OCT 09 10:57:57 | 896 | 450 | -0.13 |
| 2022 OCT 09 11:37:46 | 908 | 450 | -0.03 |
| 2022 OCT 09 11:46:47 | 896 | 450 | -0.00 |
| 2022 OCT 09 11:56:47 | 896 | 450 | 0.08 |
| 2022 OCT 09 12:05:46 | 862 | 450 | 0.08 |

Extended Data Table 6: Goldstone radar Doppler measurements of Dimorphos relative to Didymos. The fourth column shows the post-fit residuals (observed - computed) for the solution in Table 3, normalized by the 1-sigma uncertainty listed in the third column.

| **Receive time (UTC)** | **Doppler (Hz)** | **Unc. (Hz)** | **Residuals (sigma)** |
|---|---|---|---|
| 2022 SEP 27 11:22:02 | -3.00 | 2.00 | 0.12 |
| 2022 SEP 27 11:49:09 | -5.00 | 2.00 | -0.22 |
| 2022 SEP 28 10:23:24 | -4.00 | 2.00 | 0.23 |
| 2022 SEP 30 10:22:13 | -6.00 | 2.00 | -0.32 |
| 2022 OCT 01 10:05:51 | -2.50 | 2.00 | -0.27 |
| 2022 OCT 02 11:04:28 | 5.00 | 2.00 | -0.54 |
| 2022 OCT 04 09:58:15 | 7.00 | 2.00 | -0.02 |
| 2022 OCT 06 12:44:16 | -8.00 | 2.00 | -0.18 |
| 2022 OCT 06 12:57:45 | -8.00 | 2.00 | -0.33 |
| 2022 OCT 12 09:37:43 | 8.00 | 2.00 | -0.33 |
| 2022 OCT 12 10:26:49 | 9.00 | 2.00 | 0.20 |
| 2022 OCT 13 09:44:09 | 7.00 | 2.00 | -0.27 |

Extended Data Table 7: Didymos-relative optical astrometry of Dimorphos.

| Time (UTC) | ΔRA (deg) | ΔRA unc. (deg) | ΔRA residual (sigma) | ΔDEC (deg) | ΔDEC unc. (deg) | ΔDEC residual (sigma) |
|---|---|---|---|---|---|---|
| 2022-09-26 23:10:58.235 | -0.0514196 | 0.0038304 | 0.202 | -0.0125218 | 0.0034673 | 0.08 |
| 2022-09-26 23:11:04.975 | -0.0534928 | 0.0039585 | 0.12 | -0.0117131 | 0.0034132 | 0.45 |
| 2022-09-26 23:11:11.715 | -0.055801 | 0.0040724 | 0.016 | -0.0125985 | 0.0034639 | 0.32 |
| 2022-09-26 23:11:18.456 | -0.0576213 | 0.004194 | 0.067 | -0.0134683 | 0.0035253 | 0.22 |
| 2022-09-26 23:11:24.233 | -0.0593477 | 0.0043055 | 0.098 | -0.0143021 | 0.0035859 | 0.11 |
| 2022-09-26 23:11:30.973 | -0.0615916 | 0.0044643 | 0.115 | -0.0140441 | 0.0035711 | 0.35 |
| 2022-09-26 23:11:37.713 | -0.0641259 | 0.0046184 | 0.11 | -0.0150902 | 0.0036472 | 0.23 |
| 2022-09-26 23:11:44.453 | -0.0667637 | 0.0047804 | 0.127 | -0.016572 | 0.003762 | 0.01 |
| 2022-09-26 23:11:51.193 | -0.069919 | 0.0049938 | 0.087 | -0.0157601 | 0.0036996 | 0.43 |
| 2022-09-26 23:11:57.933 | -0.0732115 | 0.005196 | 0.076 | -0.0170424 | 0.0038 | 0.29 |
| 2022-09-26 23:12:04.673 | -0.0766707 | 0.0054239 | 0.093 | -0.0180365 | 0.0038846 | 0.26 |
| 2022-09-26 23:12:11.413 | -0.0803118 | 0.0056719 | 0.141 | -0.019363 | 0.004004 | 0.17 |
| 2022-09-26 23:12:18.153 | -0.0849044 | 0.0059557 | 0.094 | -0.0204015 | 0.0040924 | 0.18 |
| 2022-09-26 23:12:24.893 | -0.0893027 | 0.0062723 | 0.161 | -0.0214126 | 0.0041959 | 0.22 |
| 2022-09-26 23:12:31.633 | -0.0948127 | 0.0066306 | 0.14 | -0.0225141 | 0.0042994 | 0.28 |
| 2022-09-26 23:12:39.336 | -0.102054 | 0.0070766 | 0.105 | -0.0253911 | 0.0045736 | 0.02 |